\documentclass[lettersize]{article}

\usepackage{PRIMEarxiv}

\usepackage[utf8]{inputenc} 
\usepackage[T1]{fontenc}    
\usepackage{hyperref}       
\usepackage{url}            
\usepackage{booktabs}       
\usepackage{amsfonts}       
\usepackage{nicefrac}       
\usepackage{microtype}      
\usepackage{fancyhdr}       
\usepackage{graphicx}       
\graphicspath{{media/}}     
\usepackage{amsmath}
\usepackage{graphicx}
\usepackage{pdflscape}
\usepackage{rotating} 
\usepackage{lmodern}
\usepackage{pdflscape}
\usepackage{lmodern}
\usepackage{array, multirow}

\newcommand{\tabitem}{~~\llap{\textbullet}~~}
\usepackage{afterpage}
\usepackage{capt-of}

\title{Glass-box model representation of seismic failure mode prediction for conventional RC shear walls}

\author{Zeynep Tuna Deger, Ph.D. and Gulsen Taskin Kaya, Ph.D. \\
  Istanbul Technical University \\
  \texttt{\{zeynep.tuna@itu.edu.tr, gulsen.taskin@itu.edu.tr\}} \\
}

\begin{document}
\maketitle

\begin{abstract}
The recent surge in earthquake engineering is the use of machine learning methods to develop predictive models for structural behavior. Complex black-box models are typically used for decision making to achieve high accuracy; however, as important as high accuracy, it is essential for engineers to understand how the model makes the decision and verify that the model is physically meaningful. With this motivation, this study proposes a glass-box (interpretable) classification model to predict the seismic failure mode of conventional reinforced concrete shear (structural) walls. Reported experimental damage information of 176 conventional shear walls tested under reverse cyclic loading were designated as class-types, whereas key design properties (e.g. compressive strength of concrete, axial load ratio, and web reinforcement ratio) of shear walls were used as the basic classification features. The trade-off between model complexity and model interpretability was discussed using eight Machine Learning (ML) methods. The results showed that the Decision Tree method was a more convenient classifier with higher interpretability with a high classification accuracy than its counterparts. Also, to enhance the practicality of the model, a feature reduction was conducted to reduce the complexity of the proposed classifier with higher classification performance, and the most relevant features were identified, namely: compressive strength of concrete, wall aspect ratio, transverse boundary,  and web reinforcement ratio. The ability of the final DT model to predict the failure modes was validated with a classification rate of around 90\%. The proposed model aims to provide engineers interpretable, robust, and rapid prediction in seismic performance assessment. 
\end{abstract}


\section{Introduction}
\label{intro}
Reinforced concrete buildings constitute the majority of the existing stock in earthquake-prone countries and include piers, walls, and/or columns to resist lateral loads due to earthquake ground shaking \cite{Wasti2006,Verderame2009}. Seismic behavior (strength and ductility) and failure mechanism of such vertical structural elements are influenced by geometry and support conditions of the section, mechanical properties of structural materials (i.e., concrete and steel), amount and detailing of transverse and longitudinal reinforcement, as well as the loading pattern. Failure mechanisms of shear walls and columns are analogous and are typically classified as shear-controlled and flexure-controlled, identified by reaching shear strength prior to yielding of longitudinal reinforcement, and yielding in flexure prior to reaching the shear strength, respectively. Structural elements showing shear-flexure interaction yield in flexure before reaching the shear strength with a reduction in nonlinear deformation capacity due to the presence of shear. 
	
\begin{figure}
		\centering
		\includegraphics[width=0.7\columnwidth]{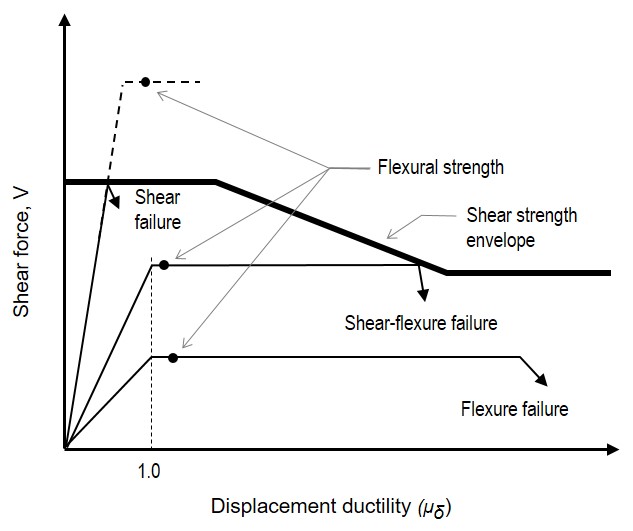}
		\caption{Conceptual shear strength model \cite{Priestley1994,Zhu2007}}
		\label{fig:envelope}
	\end{figure}

	Shear-controlled walls and columns exhibit rapid strength loss with little-to-no nonlinear deformation capacity (or ductility), i.e., brittle behavior, whereas flexure-controlled members are able to attain large inelastic flexural deformations, i.e., ductile behavior. Shear failure, flexure failure, and shear-flexure interaction can be clearly observed when shear and flexural behaviors are assessed together, as made by Priestley et al. \cite{Priestley1994} and Zhu et al. \cite{Zhu2007} for reinforced concrete columns based on a conceptual shear strength model (Fig. \ref{fig:envelope}). Fig.\ref{fig:envelope} indicates that the shear strength degradation is assumed linear, whereas failure is identified by comparing flexural strength to shear strength. Brittle shear failure is expected if flexural strength is greater than shear strength, whereas ductile flexure failure is ensured if flexural strength is smaller than the residual shear strength. For columns with shear-flexure interaction, failure is assumed where flexural strength intersects with the shear strength envelope. 
	
		Accurate prediction of the failure modes of structural elements is essential to achieve a realistic assessment of inelastic response and accurate risk assessment of existing buildings. ASCE 41-17 \cite{american2017asce} (in accordance with ACI Committee 369 \cite{ACI369}) provides classification criteria for columns based on the ratio of lateral load capacity (shear strength) to the shear demand associated with reaching the nominal flexural capacity along with the transverse reinforcement detailing. The classification of failure modes in columns was investigated by several researchers (e.g.\cite{Alcantara2000}, \cite{yoshikawa2001ductility}, and \cite{Zhu2007}). Alcantara and Imai \cite{Alcantara2000} classified column failure modes in three categories, namely: shear, bond splitting, and flexure, based on strain distributions in the longitudinal reinforcement. Yoshikawa and Miyagi \cite{yoshikawa2001ductility} suggested the use of the ratio of shear strength to flexural strength to classify columns in three failure modes (shear, shear-flexure, and flexure). Zhu et al.\cite{Zhu2007} classified columns in two failure modes (shear and flexure) using the following three wall properties: transverse reinforcement ratio, aspect ratio, and the ratio of plastic shear demand to shear strength.
	
	The shear walls are typically designed to show ductile behavior with adequate reinforcement and proper detailing. However, experimental results demonstrated that shear wall behavior depends on the wall characteristics such that squat (short) wall responses are governed by shear, and taller (slender) wall responses are dominated by flexural behavior \cite{Massone2004,Orakcal2009}. The squat walls, typically defined as the walls with an aspect ratio ($h_w/l_w$) less than 1.5, are classified as the shear-controlled, whereas the slender walls, with an aspect the ratio greater than 3.0, are identified as the flexure-controlled (ASCE 41-17). It is worth noting that the squat walls may show the flexure-controlled behavior with the adequate design and the reinforcement detailing \cite{Salonikios1999}, whereas the slender walls may have  shear (web crushing) failure \cite{sittipunt2001cyclic} with poor reinforcement and confinement. 

	The classification of the shear walls has been studied by various researchers based on  experimental data. Wood \cite{wood1991} classified wall failure modes as either the flexure or the shear based on the experimental evidence of 27 wall specimens, and classification was predicated on the wall shear stress level ($4\sqrt{f_c}$). Fang \cite{fang1992} classified the failure modes based on 45 relatively tall specimens in four categories, namely: the shear failure, the flexure failure, the flexural-shear, the failure, and the out-of-plane failure. The predicted failure mode was identified based on the shear span ratio, the ratio of shear strength to the flexural strength, the shear demand, and the wall detailing. Greifenhagen \cite{greifenhagen2005static} suggested a classification criterion for low-rise shear walls (shear span ratio $\leq 1.0$) based on shear demand-to-capacity ratio, and the shear span ratio; where failure was identified as elastic shear, low-ductile to moderate-ductile response, and brittle shear. Abbasnia \cite{abbasnia2011new} suggested a decision tree procedure based on 105 specimens to classify the shear, the flexure, and the flexural-shear failure, including web crushing and sliding failure. More recently, Grammatikou \cite{grammatikou2015} proposed a predictive model to determine failure modes in the following categories: flexure, diagonal tension, diagonal compression, and sliding shear using 866 shear wall specimens.

 	 As prediction of the failure modes is essentially a classification problem, the use of machine learning methods was considered in this study.   Recently, the ML methods have received a lot of attention in the field of structural earthquake engineering due to their easy implementation and capability to model the input-output relations via clustering, classification, or regression \cite{Sudret2014b,Jiang2015a}.  For example, Bourinet addressed the estimation of small failure probabilities, which is a very challenging task in structural reliability, by subset simulation and SVM classifier in combined with active learning \cite{Bourinet2011a}. Salazar assessed the potential of some state-of-art machine learning methods for dam modeling in terms of displacements and leakage as an alternative to deterministic models \cite{Salazar2015}, whereas Zhang proposed a ML-based framework to assess post-earthquake structural safety by introducing the concept of response and damage patterns for any damaged building \cite{Zhang2018}. More recently, the utility of the ML methods has been investigated to predict earthquake-induced building damage \cite{Mangalathu2020classifying}, classification of failure modes for reinforced masonry shear walls \cite{Siam2019}, frame structures with masonry infill panels  \cite{HuangBurton2019}, reinforced concrete beam-column joints, and shear walls \cite{Mangalathu2018,mangalathu2020data}. The proposed models in literature are generally highly nonlinear complex models.
 	 
 Literature review shows that ML methods are very convenient tools on many different applications related to  earthquake engineering and seismic risk assessment as they demonstrate promising and encouraging results.  The proposed ML classification models in the literature are reasonably accurate but are not transparent from the point of complete understanding by users, i.e., they are black-box models. In machine learning,  \textit{a black-box model is a system that does not reveal its internal mechanisms} \cite{molnar2019interpretable}.  From the engineering point of view, the main concern on the black-box models is that the model might fail in unexpected cases due to some possible inconsistency between the physical model and its machine learning representation while providing very good performance \cite{RibeiroSG16}. The literature of the interpretation of machine learning algorithms, also known as Explainable Artificial Intelligence (xAI) in Computer Science, is quite extensive even though this concept is more recently exploited. The interpretation is addressed with two main branches: \textit{transparent models} and \textit{post-hoc explainability}  \cite{BarredoArrieta2020}. The former refers to a machine learning model which is understandable for a human \cite{Lipton2018}, and the latter includes the post-hoc type methods that are  specifically designed for explaining internal structure of the black-box type learning models \cite{Robniksikonja2008589}.  Transparency and interpretability are two concepts that are strongly tied to each other. The examples of the transparent models are Logistic Regression, Decision Trees (DT), k-Nearest Neighbors, Rule-base Learners, General Additive Models, and Bayesian methods. Among them, the DT easily fulfil most of the constraints for the transparency in terms of providing a  decomposable  and algorithmic transparent model  at the expense of low generalization capability \cite{QUINLAN1987221}. Therefore,  the trade-off between the performance of a model and its transparency should be properly established by means of  feature relevance techniques \cite{ZHANG2019158}. Besides, it is important for engineers to understand how the model makes the decision and verify that the model is physically meaningful. The technical challenge of explaining decisions made by the models is referred to as the interpretability problem. Therefore,  explanatory models, known as glass-boxes, are more useful than the black-box algorithms since they provide necessary and sufficient information to assess how the input variables of a system depend on the system output \cite{Bastian2018}.  The disadvantages of black-box models and the importance of interpretability in machine learning have been pointed out by various researchers \cite{pena2018opening,ExplainingExplanationsMIT2018,murdoch2019interpretable,Schmidt2019}, yet, not in the field of earthquake engineering. 
 As summarized in the literature review, the trade-off between model complexity and model interpretability is typically disregarded in the earthquake engineering field despite the importance of the explanatory models. This study aims to fill this gap and assess the ML methods to provide an interpretable (glass-box) model for failure mode classification of conventional reinforced concrete shear walls while ensuring reliability and high accuracy. To achieve this, eight ML methods are utilized to classify three wall failure modes based on the experimental evidence such that the highest possible accuracy is achieved by ensuring physical significance along with a high interpretable model having the least number of features possible. To ensure such a model, the dimensionality of the problem is reduced with a feature selection algorithm. Features that are not physically meaningful are removed from the learning model even though they improve the classification performance. The class-wise and overall classification accuracy of the failure modes were compared for the considered ML classification methods along with the ASCE 41-17 classification criteria. Novel aspects of this study are i) investigating the state-of-the-art machine learning methods from the Explainable Artificial Intelligence (xAI) point of view, ii) establishing a convenient trade-off between interpretability and classification performance, and iii) obtaining a robust, interpretable, and physically meaningful classification model without renouncing the high accuracy accuracy for predicting failure mode for conventional reinforced concrete shear walls. The proposed classification model is aimed to provide a preliminary prediction for the shear wall to achieve better nonlinear modeling for a detailed performance assessment.

\section{Shear Wall Test Database}
\label{sec:1}
A total of 176 shear wall specimens from 19 different experimental studies conducted worldwide were collected based on a review of available research. The database consists of conventional reinforced concrete shear walls that would typify the existing building stock. It is noted that similar databases are available in the literature, \cite{grammatikou2015,gulec2011,mangalathu2020data,NEESHub}, consisting of a much wider range of data; however, they were not suitable for the use of this study directly as the following walls do not represent conventional walls in the existing building stock:
	(i) specimens that were constructed using high strength material ($>$40 MPa and $>$500 MPa for concrete and reinforcing steel, respectively),
	(ii) those that were constructed with diagonal web reinforcement or with no vertical and/or horizontal web reinforcement, 
	(iii) walls with openings or hollow sections,
	(iv) walls that were constructed using composite materials, 
	(v) specimens  tested under monotonic loading (as the focus is earthquake-induced damage),
	(vi) repaired or strengthened walls. 
	
\begin{figure}
		\centering
		\includegraphics[width=\linewidth]{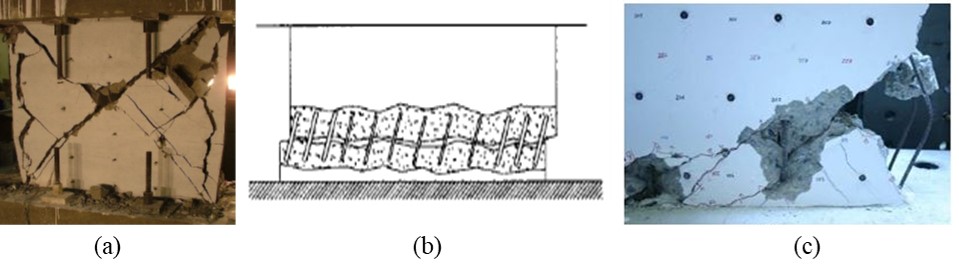}
		\caption{(a) Diagonal web crushing \cite{massone2006rc}, (b) Sliding shear failure \cite{salonikios2007analytical}, (c) Buckling of longitudinal boundary reinforcement \cite{greifenhagen2005static}}
		\label{fig:resimler}
	\end{figure}

		Failure modes of shear wall specimens can be categorized based on two approaches. One approach is theoretical and depends on the shear and the flexural strength values of the shear wall, i.e. the ratio of the shear force developed to the shear force associated with reaching the wall nominal flexural capacity ($V/V@M_n$), similar to the recommendations of ASCE 41. Another approach is to assign failure modes with respect to the experimental evidence (reported failure modes), based on the criteria summarized in Table \ref{table:failuremodes}. In this study, failure mode class assignments were made based on the latter approach as they represent ground truth (actual class labels).  The ASCE 41 classification criteria are also discussed in detail in the following sections.  The failure types for shear-controlled walls include diagonal tension, web crushing, and sliding; whereas those for flexure-controlled walls include concrete crushing and reinforcement buckling at the wall boundary. The failure types observed in the shear-flexure walls commonly consist of shear failure followed by concrete spalling and crushing at wall boundaries or sliding shear failure after yielding of longitudinal boundary reinforcement. Three of the most common damage occurred during the tests, including web crushing, sliding shear after yielding of boundary reinforcement, and buckling of longitudinal boundary reinforcement, are depicted in Fig. \ref{fig:resimler}. The database includes 50 specimens for the shear-controlled walls, 57 specimens for the flexure-shear walls, and 69 specimens for the flexure-controlled walls. 
	
\begin{table}[!h]
\centering
		\small
		\caption{Reported failure modes of the walls.}
		\label{table:failuremodes}
		\begin{tabular}{llll}
			\hline
			& \begin{tabular}[c]{@{}c@{}}\textbf{Shear}\\   \textbf{failure}\end{tabular}                                       & \begin{tabular}[c]{@{}c@{}}\textbf{Flexure-shear failure}\\   \textbf{}\end{tabular}                                       & \begin{tabular}[c]{@{}c@{}}\textbf{Flexure failure}\\   \textbf{}\end{tabular}                                \\ \hline                                                                  
			\begin{tabular}[c]{@{}l@{}}\textbf{Expected} \\ \textbf{failure} \\ \textbf{types}\end{tabular} & \begin{tabular}[c]{@{}l@{}} diagonal\\   tension\\ failure\\   
				\\ web\\   crushing\end{tabular} & \begin{tabular}[c]{@{}l@{}}  
				\\ shear failure followed \\ by concrete spalling and \\ crushing at wall boundaries\\   
				\\ sliding shear failure after \\ yielding of longitudinal \\ boundary reinforcement\end{tabular} & \begin{tabular}[c]{@{}l@{}} concrete spalling\\    and crushing\\  
				\\ rebar buckling and \\ lateral instabilities \\ at boundary elements\end{tabular}
			\\ \hline
		\end{tabular}
	\end{table}
	
\begin{figure}[!h]
		\centering
		\includegraphics[width=1\columnwidth]{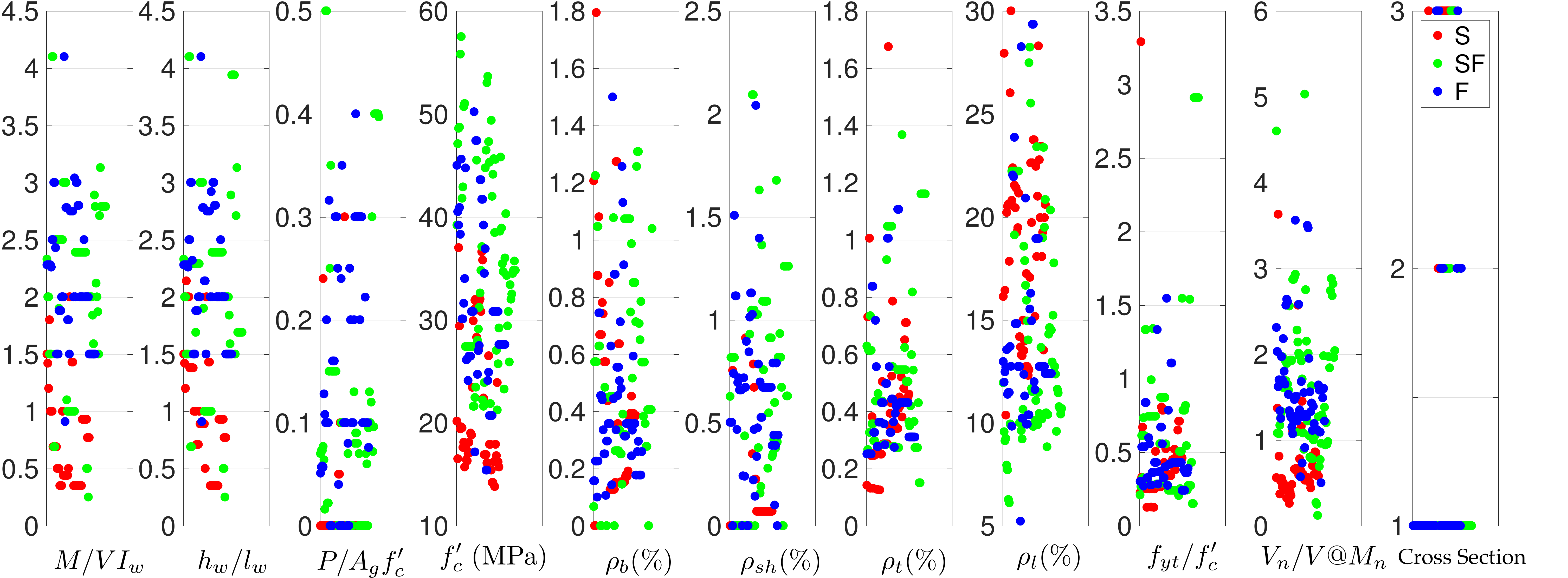} 
		\caption{Distribution of key wall properties (176 specimens)}
		\label{fig:dataScatter}
	\end{figure}

	Key wall properties include specimen geometry (rectangular, barbell, and flanged), wall aspect ratio ($h_w/l_w$) and shear span ratio ($M/Vl_w$), mechanical properties of concrete and steel ($f'_c$ and $f_{yt}/f'_c$), transverse and longitudinal boundary reinforcement ratios ($\rho_{sh}$ and $\rho_{b}$, respectively), transverse and longitudinal web reinforcement ratios ($\rho_{t}$ and $\rho_{l}$, respectively), as well as axial load ratio ($P/A_gf'_c$) and shear strength-to-shear at flexural strength ratio ($V_n/V@M_n$). The distribution of key wall design properties are presented in Fig. \ref{fig:dataScatter} for each failure mode. Further details on the database can be obtained elsewhere. \cite{deger2019empirical}.
	
		The classification criteria provided by ASCE 41-17 were assessed using the experimental database. For this purpose, classification criteria suggested for shear walls (ASCE-W), as well as that recommended for columns (ASCE-C) were applied to the shear walls in the database. When the ASCE-C is used, the column transverse reinforcement details are assumed to be associated with the wall boundary elements. As required by the ASCE 41-17, boundary reinforcement was verified to satisfy the following conditions to qualify for flexure-controlled category: $A_v/b_ws \geq 0.002$ and $s/d_o\leq 0.5$. Table \ref{table:ascecriteria} summarizes the ASCE 41-17 classification criteria, whereas Table \ref{table:databasesummary} tabulates the associated predictions of the three failure modes. 
	
\begin{table}[!h]
		\small
		\caption{Classification of failure modes based on ASCE 41-17.}
		\begin{tabular}{lllllll}
			
			\hline
			\multicolumn{2}{l}{\textbf{Wall criteria}}                                                                                   
			& \multicolumn{4}{l}{\textbf{Column criteria adapted to walls}}                      
			\\ \hline
			\begin{tabular}[l]{@{}l@{}}Aspect\\   Ratio\end{tabular} & \begin{tabular}[c]{@{}l@{}}Failure\\   Mode\end{tabular} &
			\begin{tabular}[l]{@{}l@{}}Shear\\   Capacity\\ Ratio\end{tabular} & \begin{tabular}[l]{@{}l@{}}ACI\\   conforming\\ hoops with\\ 135-degree\\ hooks\end{tabular} & \begin{tabular}[l]{@{}l@{}}Non-conforming\\   hoops with\\  135-degree\\ hooks\end{tabular} & \begin{tabular}[l]{@{}l@{}}Closed\\   hoops with\\ 90-degree\\ hooks\end{tabular} & Other \\ \hline
			$\dfrac{h_w}{l_w}<1.5$       &  S 
			& $\dfrac{V_n}{V@M_n}\geq1.67$    & F   & SF  & SF  & SF    \\ 
			$1.5\leq\dfrac{h_w}{l_w}\leq3.0$  & SF
			&$1.0\leq\dfrac{V_n}{V@M_n}<1.67$ & SF  & SF  & SF  & S     \\
			$\dfrac{h_w}{l_w}>3.0$      	  & F   
			&$\dfrac{V_n}{V@M_n}>1.0$         & S   & S   & S   & S    
			\\ \hline
		\end{tabular}
	\label{table:ascecriteria}
	\end{table}

	\begin{landscape}
	\begin{table}[]
	\centering
			\tiny
			\caption{Overview of the experimental database.}
			\label{table:databasesummary}
			\begin{tabular}{cccccccccccccc}
				\hline
	\textbf{\#} & \textbf{$M/Vl_w$} & \textbf{$h_w/l_w$} & \textbf{ $P/A_gf'_c$} & \textbf{$f'_c$ (MPa)} & \textbf{ $\rho_{b}$ (\%)} & \textbf{ $\rho_{sh}$ (\%)} & \textbf{ $\rho_{t}$ (\%)} & \textbf{ $\rho_{l}$ (\%)} & \textbf{$f_{yt}/f'_c$}  & \textbf{X-sec*} & Failure**    \\
				\hline
 2           & 2.78           & 2.78           & 0.0             & 47.4               & 0.44 - 1.5       & 0.84 - 1.01       & 0.50             & 0.37             & 10.2       & R     & \textbf{F}        \\
 1           & 3.04           & 2.92           & 0.1             & 34.5               & 0.48             & 0.48              & 0.27             & 0.27             & 16.3      & B  & \textbf{F}        \\
3           & 2.50           & 2.29           & 0.15 - 0.35     & 27.4               & 0.45             & 0.00              & 0.44             & 0.74             & 22.2      & R  & \textbf{SF}       \\
1           & 1.50           & 1.50           & 0.0         & 20.2      & 1.21  & 0.00              & 0.14             & 0.23            & 16.1            & R          & \textbf{S}        \\
4           & 0.25 - 1.0     & 0.25 - 1.0     & 0.0             & 21.2 - 27          & 0.85           & 0.91      & 0.50             & 0.25 - 0.51      & 19.0 - 23.4     & F   &\textbf{S, SF}    \\
6  & 2.26-2.28      & 2.26-2.28      & 0.05 - 0.13     & 38.3-45.6          & 0.1 - 0.23       & 0 - 1.51          & 0.25             & 0.27-0.54        & 11.4 - 13.5   & R   & \textbf{F}        \\
2           & 0.69           & 0.69           & 0.0         & 51                 & 0.00             & 0.00              & 0.30             & 0.30             & 9.9     & R         & \textbf{SF}       \\
14          & 0.35 - 1       & 0.71 - 2.0     & 0.0             & 15.7 - 19.6        & 0.44 - 1.08 & 0.00              & 0.13 - 0.38      & 0.13 - 0.26      & 17.8 - 30  & R       & \textbf{S}        \\
6          & 1.9 - 2.5      & 1.69 - 2.29    & 0.2      & 27.4   & 0 - 0.48         & 0 - 0.44          & 0.44 - 0.46      & 0.74 - 1.34      & 19.1 - 22.2      & R              & \textbf{SF}       \\
2      & 3.00           & 3.00           & 0.2 - 0.25      & 15.4   & 0.91 - 1.13      & 0.7 - 1.4         & 0.42             & 0.42     & 29.4                  & R,B           & \textbf{F}        \\
2      & 2.80           & 2.80           & 0.1      & 24.1 - 24.9       & 0.31             & 0.53              & 1.11             & 1.11        & R                       & \textbf{F}        \\
19       & 0.35 - 0.93    & 0.35           & 0.0    & 13.8 - 26.5     & 0.15 - 0.45      & 0.07              & 0.34 - 0.71      & 0.34 - 0.71      & 13.5 - 28.3   & F          & \textbf{S}        \\
6           & 1.00           & 1.00           & 0.1             & 28 - 37.1     & 0 - 0.26         & 0 - 2.1           & 1.05             & 0.87             & 14 - 18.6        & R    & \textbf{SF}       \\
2           & 2.00           & 2.00   & 0.0    & 29.2 - 39.2        & 0.71       & 0.34              & 0.36             & 1.55     & 15.5 - 20.8    & R                       & \textbf{SF, F}    \\
1           & 1.10           & 1.90           & 0.0         & 45.5         & 1.08             & 0.93              & 0.93             & 0.56     & 10.6     & R                       & \textbf{SF}       \\
4           & 1.87 - 2.79    & 3.94           & 0.1 - 0.13      & 36 - 40.3     & 0.38 - 0.57 & 0.82 - 0.93       & 0.27             & 0.27 - 1.54      & 13.0 - 17.8     & R     & \textbf{SF}       \\
1           & 2.43           & 2.32           & 0.3             & 44.7               & 0.44      & 0.72              & 0.72             & 0.60             & 9.8    & R         & \textbf{F}        \\
6           & 0.44           & 0.89           & 0.0 - 0.1       & 28.3 - 32     & 0.13             & 0.00        & 0.29             & 0.26        & 13.3 - 15.0     & R       & \textbf{S}        \\
1           & 2.33           & 2.33           & 0.1         & 39.2       & 0.07     & 0.63             & 0.63             & 0.21             & 11.9      & R                       & \textbf{SF} \\
13          & 2.39           & 2.39           & 0.0 - 0.13      & 21.9 - 53.6        & 0.15 - 1.07      & 0.16 - 1.63       & 0.27 - 1.37      & 0.21 - 0.24      & 9.8 - 23.4  & R, B, F                 & \textbf{SF}       \\
8           & 1.50           & 1.50           & 0.08 - 0.22      & 27.6               & 0.18 - 0.3       & 0.1 - 0.79        & 0.31             & 0.24 - 0.39      & 12.4     & R,B,F                   & \textbf{F}        \\
2           & 2.71 - 3.13    & 2.71 - 3.13    & 0.07 - 0.1      & 25.9 - 33.4        & 0.28       & 0.00              & 0.15             & 0.15             & 9.6 - 12.4  & R    & \textbf{SF}       \\
1           & 0.50           & 0.50           & 0.0             & 27.2               & 0.15       & 1.13              & 1.67             & 0.81             & 14.0   & R           & \textbf{S}        \\
6           & 2.00           & 2.00  & 0.0   & 31.8 - 45.8     & 1.26 - 1.31      & 0.35 - 0.79       & 0.35 - 0.6       & 0.79          & 8.8 - 17.2      & R          & \textbf{S, SF, F} \\
1           & 0.91           & 0.91           & 0.0             & 17.2               & 0.14             & 0.00              & 0.28             & 0.28             & 28.3             & R                       & \textbf{F}        \\
3           & 1.50           & 1.50           & 0.0             & 24.1 - 26.2        & 0.25 - 0.33 & 0.66              & 0.28 - 0.56      & 0.28 - 0.56      & 21.9 - 23.9      & R     & \textbf{F} \\
4     & 1 - 1.5        & 1 - 1.5        & 0.0 - 0.07      & 21.6 - 27.5   & 0.25 - 0.35      & 1.03         & 0.28 - 0.56      & 0.28 - 0.56      & 20.9 - 28.2    & R &\textbf{SF, F}    \\
1    & 2.89           & 2.89           & 0.0             & 23.3               & 0.78             & 0.41              & 0.36             & 0.24             & 20.3       & R  & \textbf{SF}       \\
2           & 1.43           & 1.43           & 0.0             & 35.8 - 36.6        & 0.64      & 0.23     & 0.52 - 0.79      & 0.39 - 0.52      & 12.3 - 12.6        & B & \textbf{S}        \\
3     & 4.10           & 4.10           & 0.25 - 0.50     & 41.8 - 50.2        & 0.63    & 0.00              & 0.36             & 1.33             & 5.2 - 6.3  & R         & \textbf{SF, F}    \\
4   & 3.00           & 3.00           & 0.0             & 21.6 - 23.5        & 0.45             & 0.00              & 0.28             & 0.28             & 9.2 - 10     & R & \textbf{SF}       \\
2   & 3.00           & 3.00           & 0.1             & 31.6 - 34          & 0.46             & 0.47 - 0.7        & 0.33             & 0.33             & 12.7 - 13.7       & R                     & \textbf{F}        \\
4    & 2.75           & 2.75           & 0.1             & 41.7 - 43.6        & 0.46 - 0.55      & 0.14 - 2.04       & 0.33 - 0.44      & 0.326 - 0.4      & 9.9 - 10.4  & F    & \textbf{F}        \\		
2           & 1.2 - 1.42     & 1.2 - 1.42     & 0.0             & 16.5 - 37          & 0.00             & 0.00              & 0.26 - 0.73      & 0.33 - 3.29      & 27.9     & R    & \textbf{S}    \\
6           & 2.79           & 1.69           & 0.4             & 32 - 35.7   & 0 - 1.04         & 0.63 - 1.26       & 1.16             & 2.91             & 10.4 - 11.6       & R      & \textbf{SF}  \\
5   & 1.5-2.0        & 1.5-2.0        & 0.02 - 0.08     & 47.1 - 57.5        & 0.57 - 1.22      & 0.82              & 0.27 - 0.73      & 0.27 - 0.73      & 7.7 - 9.6 & R    & \textbf{SF}       \\
2           & 1.59 - 1.84    & 1.59 - 1.84    & 0.06 - 0.07     & 35    & 0.65 - 0.71      & 0.38 - 1.68       & 0.54 - 0.82      & 0.54 - 0.82      & 14.3 - 14.6    & R,B & \textbf{SF}       \\
13   & 1.5 - 2.0      & 2.00           & 0.1 - 0.4       & 20.7 - 30.8        & 0.24 - 0.59      & 0.67       & 0.43             & 0.43             & 12.7 - 18.9  & R       & \textbf{S, SF, F} \\
6    & 1.88 - 2.0     & 1.88 - 2.0     & 0.16 - 0.30     & 26.5 - 30.8        & 0.11 - 0.36  & 0 - 0.9  & 0.36 - 0.43      & 0.36 - 0.43      & 12.7 - 14.8   & R   & \textbf{F}        \\
3           & 1.80           & 2.14           & 0.24 - 0.35     & 24.7 - 29.4        & 0.88 - 1.79  & 0.75 - 1.13      & 1.00       & 0.67             & 12.3      & R      & \textbf{S, F}     \\
2   & 2.50           & 2.50           & 0.10 - 0.20     & 30.1    & 0.74             & 1.12              & 0.84             & 0.84             & 11.5     & R       & \textbf{F}        \\
				
				\hline
				\multicolumn{14}{l}{* Wall cross sections: R = rectangular, B = barbelled, F = flanged; **Failure types: S = shear-controlled, SF = shear-flexure, F = flexure-controlled}                                                                                             \\ 
			\end{tabular}
	     	\end{table}
			\end{landscape}

	$V@M_n$  is the shear force in the wall associated with reaching the wall nominal moment capacity ($M_n$) for the given distribution of lateral forces used in the test. Wall nominal shear strength   can be obtained as shown in Eq. (\ref{eq:eq1}) by using expected (or tested) material properties, 
		\begin{equation}
	V_n=A_{cv} (\alpha_c \lambda \sqrt{f'_c}+\rho_t f_{yt} )
	\label{eq:eq1}
	\end{equation}
	where $A_{cv}$ is the wall web length multiplied by the web thickness in the direction of the applied shear force; $\alpha_c$ is a coefficient to account for aspect ratio, calculated by $2\leq(\alpha_c=6-2h_w/l_w)\leq3$, where $h_w$ and $l_w$ are wall height and length, respectively. $f'_c$  is the specified concrete strength; whereas $\rho_t$ and $f_{yt}$ are the reinforcement ratio and yield strength of the web horizontal reinforcement, respectively.

	\section{Implementation of Machine Learning Techniques and Results}
	
	The key shear wall properties (Fig.\ref{fig:dataScatter}) and the  reported failure modes (S, SF, and F) were designated as input variables (features) and output variables (class labels), respectively. The classifications were conducted using all available (total of 11) features. In all implementations, training and test datasets were randomly drawn by 10-fold cross validation and are formed as 90\% and 10\% of the entire database, respectively. The performance of the methods were evaluated based on overall classification accuracy (median and mean)  over ten trials along with sensitivity and precision  calculated based on the confusion matrix associated with the classification method.  In tuning the parameters of the machine learning methods, 5-fold cross validation approach was used. For SVM, the potential parameter combinations of the RBF kernel parameters and polynomial kernel parameters were tested in a user defined range, and the combinations providing the highest classification accuracy on the validation set were used in generating the learning model of corresponding SVM classifier. Since SVM is originally designed as a binary classifier, the one-against-all strategy was  considered for the multi-class case. 	The number of trees and the number of features for the best splitting when growing the trees for Random Forest were set to 200 and to the square root of the number of input features, respectively, as usually recommended in the literature \cite{Izenman}. 	The GP classifier enables an automatic way of tuning the kernel parameters by using gradient-based optimization routines to maximize the marginal likelihood. All  of the experiments were implemented in Matlab and run on a personal computer with Dual Core Intel Core i7 CPU (3.50GHz), 16GB RAM and MacOS operating system.

\begin{figure}[!h]
		\centering
		\includegraphics[width=1.005\columnwidth]{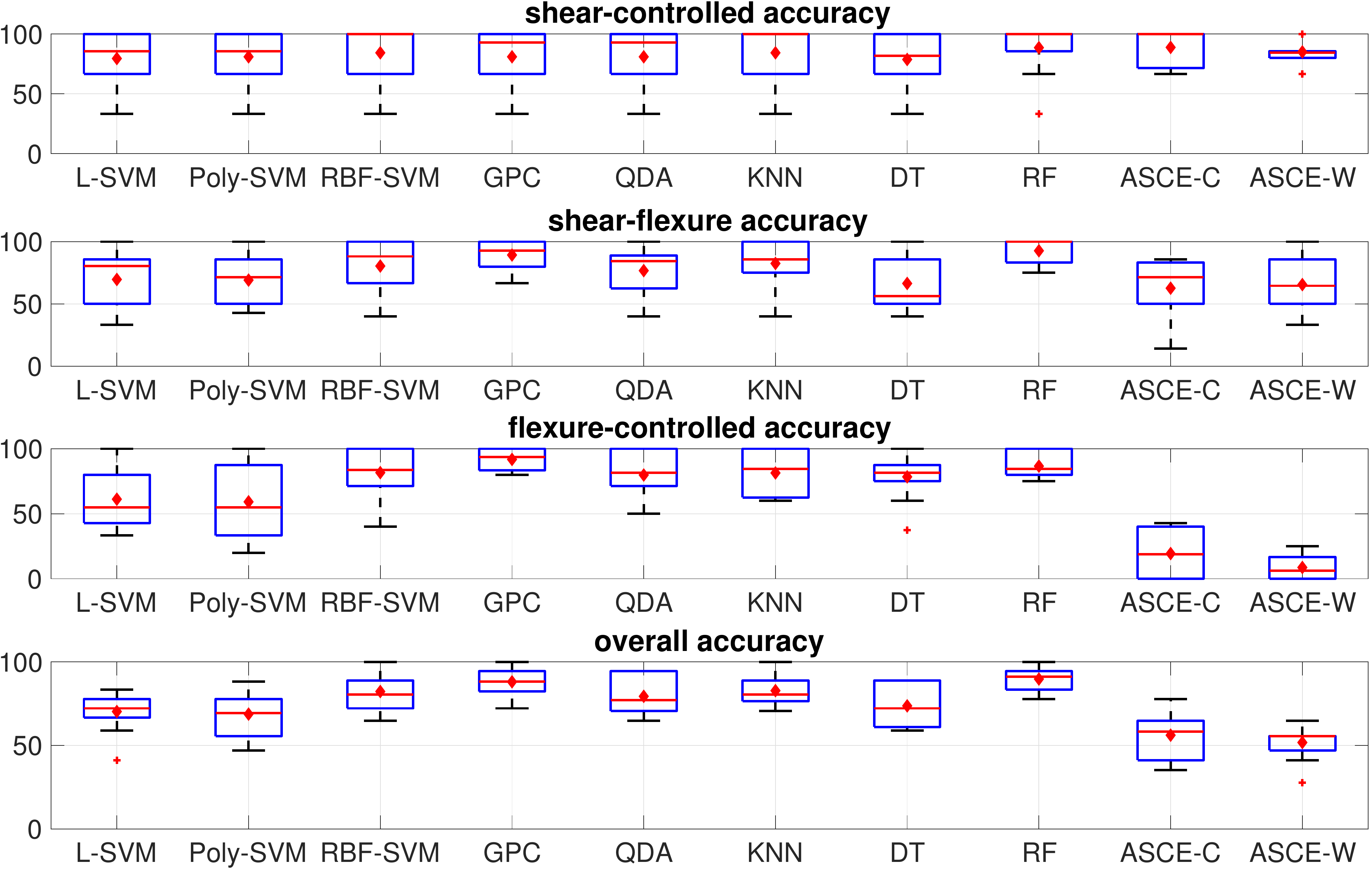} 
		\caption{Classification accuracy plots of all the methods with 11 features on 10-fold cross validation. The red diamond shows the mean accuracy.}
		\label{fig:case1_run1}
	\end{figure}

The class-wise, overall classification accuracy, and  Kappa coefficients \cite{Cohen1960} of the failure modes using eight classification methods along with the ASCE 41-17 classification criteria are summarized in Fig. \ref{fig:case1_run1} and Table \ref{table:case1_run1}. 
 According to the Cohen’s Kappa statistic interpretation, slight to moderate agreement is achieved with ASCE-C and ASCE-W, whereas moderate to substantial agreement is achieved with  ML methods, meaning that ML methods perform better than classical approaches. The classification accuracies achieved by the eight ML methods varied between 80\% and 90\% for the shear failure mode (S), whereas the highest accuracy was achieved by the ASCE-C classification criteria. The ML methods made predictions with relatively more disperse accuracies (varied from 60\% to 92\%) for the SF and the F failure modes, all of which were able to predict with higher accuracy than the ASCE-C and ASCE-W criteria. It should be noted that the ASCE 41 is a seismic evaluation and assessment standard; therefore, the predictions are rather on the brittle side to ensure conservative assessment. This is the reason why the prediction accuracy of ASCE criteria is lower for the SF and the F failure modes despite the high accuracy obtained for the S failure mode.
	
		\begin{table}[!h]
		\centering
		\footnotesize
		\caption{Class-wise mean accuracies along with the overall accuracies based on 10-fold cross validation experiments.}
		\label{table:case1_run1}
		\begin{tabular}{p{0.8cm}p{0.8cm}p{0.8cm}p{0.8cm}p{0.8cm}p{0.8cm}p{0.8cm}p{0.8cm}p{0.8cm}p{0.9cm}p{0.9cm}}
			\cline{2-11}
			& \textbf{L-SVM} & \textbf{Poly-SVM} & \textbf{RBF-SVM} & \textbf{GPC} & \textbf{QDA} & \textbf{KNN} & \textbf{DT} & \textbf{RF} & \textbf{ASCE-C} & \textbf{ASCE-W} \\ \cline{1-11}
			\textbf{S}  & 79.5           & 81.0              & 84.3             & 81.0         & 81.0         & 84.3         & 78.7        & 88.6        & 88.8            & 85.0            \\ 
			\textbf{SF} & 69.7           & 69.1              & 80.4             & 89.3         & 76.8         & 82.5         & 66.5        & 92.7        & 62.7            & 65.5            \\ 
			\textbf{F}  & 61.2           & 59.2              & 81.6             & 91.8         & 79.8         & 81.4         & 78.4        & 86.7        & 19.4            & 8.7             \\ \hline 
			\textbf{OA} & 70.3           & 68.8              & 82.3             & 88.1         & 79.4         & 82.8         & 73.7        & 89.8        & 56.1            & 51.8            \\ 
	 \textbf{Kappa} & 0.60  &	0.57 	& 0.72 	& 0.81     & 0.68    & 0.73     & 0.61  &0.84 &	0.45 &	0.39  \\ 
		\hline
		\end{tabular}
	\end{table}

 The results indicate that the classification methods implemented in this study provide better  performances than the classical approaches. However, not all of them yield explicit equations. The easy-to-use and physically meaningful equations are generally desirable for engineering practice, which was prioritized in the method selection in this study even though the higher prediction accuracies were achieved using other classification methods.  The strengths and weaknesses of all the ML methods are summarized in Table \ref{table:summary}.    Among the ML methods that yield an explicit formulation for classification (e.g. the SVM with linear and polynomial kernel, and the DT), the DT is selected as it is able to provide a single flow chart to identify all three failure modes, unlike the SVM. It is worth noting that higher prediction accuracy is achieved using the RF and the GPC, both of which provide black-box algorithms  to predict the wall failure modes.

	\clearpage
\begin{table}
\centering
			\caption{Overview of the ML methods used in this study.}
			\label{table:summary}
			\tiny
			\begin{tabular}{p{2.2cm}p{6cm}p{6cm}}
				\hline	
				\textbf{Methods} &  
				\textbf{Strengths} & 
				\textbf{Weaknesses      }        \\ \hline
				\begin{tabular}[c]{@{}l@{}} SVM with \\linear kernel and \\polynomial kernel\end{tabular} & \begin{tabular}[c]{@{}>{\textbullet\hspace{\labelsep}}l}high interpretability \\   yielding an analytical equation\\   fast to implement \\   requires less parameters to tune \\   works well with less number of samples\end{tabular}                                                                       & \begin{tabular}[c]{@{}l@{}} \tabitem low accuracy for nonlinearly distributed  data \\ \tabitem each class will have an equation, and \\  final class label  is determined based on the majority \\  voting of the predictive equations.\end{tabular}                                                                                                   \\ \hline
				SVM with RBF kernel   & \begin{tabular}[c]{@{}l@{}} \tabitem superior to SVM with linear and polynomial kernel \\   in terms of classification performance, due to its ability \\   to deal with nonlinearly distributed data via \\   nonlinear kernel function\\          \tabitem  requires more parameter to tune than linear SVM, and \\   sensitivity of the parameters has high effect on classification\\  accuracy\end{tabular} & \begin{tabular}[c]{@{}l@{}} \tabitem does not provide an explicit but an implicit \\   equation associated with a kernel matrix, \\   therefore, resulting in a low interpretability\\   \tabitem training dataset need to be retained for labeling \\   a new sample\\   \tabitem a different equation is needed for each class\\   which requires majority voting for class labeling\end{tabular} \\
				\hline
				Decision Tree                                                                            & \begin{tabular}[c]{@{}l@{}} \tabitem high interpretability \\  \tabitem provide only a single decision  graph for all the classes\\  \tabitem  probability of the class for each sample is available\\ \tabitem   can handle nonlinearly distributed dataset with a high\\   performance       \\   \tabitem  very fast for labeling the new samples\end{tabular}                                                                & \begin{tabular}[c]{@{}l@{}} \tabitem not a unique decision graph is proposed\\  \tabitem  requires a high computation during training \\ \tabitem   a parameter tuning is needed\end{tabular}                                                                                                                                                                                                           \\  \hline
				KNN                                                                                      & \begin{tabular}[c]{@{}l@{}}  \tabitem tackles with  nonlinearity of  dataset     \\   \tabitem very easy to use for labeling of a new sample\end{tabular}                                                                                                                                                                                                                                                           & \begin{tabular}[c]{@{}l@{}}\tabitem not robust to noise and dimensionality \\   of the dataset    \\  \tabitem  does not provide an equation\\  \tabitem  low interpretability\\   training set need to be retained for labeling \\   of a new sample\end{tabular}                                                                                                                                     \\ \hline
				GPR                                                                                      & \begin{tabular}[c]{@{}l@{}}  \tabitem provides high classification accuracy  \\\tabitem robust to noise on the dataset\\           \tabitem  class conditional probability is available\end{tabular}                                                                                                                                                                                                                  & \begin{tabular}[c]{@{}l@{}}\tabitem dataset is assumed to follow Gaussian distribution   \\  \tabitem  training dataset need to be retained\\ for  kernel matrix calculation  for labeling of a new sample\\   \tabitem   very low interpretability due to the type of equation\\ \end{tabular}                                                            \\ \hline
				QDA                                                                                      & \begin{tabular}[c]{@{}l@{}}\tabitem easy to implement        \\   \tabitem can handle nonlinearity of the dataset \end{tabular}                                                                                                                                                                                                                                                                                   & \begin{tabular}[c]{@{}l@{}}\tabitem low interpretability due to dependency on mean \\   vector and covariance matrix for each class \\  \tabitem mean vector and covariance matrix    need to be retained\end{tabular}                                                                                                                                                  \\ \hline
				Random Forest                                                                            & \begin{tabular}[c]{@{}l@{}}\tabitem same as the DT except that it provides higher\\   performance than the DT\end{tabular}                                                                                                                                                                                                                                                                                       & \begin{tabular}[c]{@{}l@{}}\tabitem not easy to use for labeling since many trees are\\   provided, therefore, it has low interpretability 
				\end{tabular}    
				\\  \hline                                                                                                                                                                                                                         
			\end{tabular}
	\end{table}

	Another critical issue was to design a classifier with as few features as possible  to reduce  the engineering efforts in preparing dataset. Therefore, a feature selection was conducted to lower the complexity of the classifier with the same or higher classification accuracy. Preliminary feature selection, based on Fisher Score \cite{Duda-etal01} and the SVM-RFE \cite{DBLP:journals/ml/GuyonWBV02}, indicated that aspect ratio (or shear span ratio) was the most significant feature to make a distinction between the failure modes, in accordance with the findings in previous research in literature 	 \cite{Massone2004,Orakcal2009,american2017asce}. The importance of other features in classification varied from one feature selection method to another. Therefore, a brute-force feature subset selection was conducted, that is, all physically meaningful combinations of three and four features were exhaustively evaluated and the best combination was determined based on the overall accuracy. The combinations of four features achieved considerably higher accuracy than that of three features. Therefore, feature subsets consisting of four features, one of which was either aspect ratio or shear span ratio, were considered. The most challenging drawback of the brute-force feature selection  method is high computational cost, which was insignificant for this study as the number of input features was limited and data were not too extensive. 
	
\begin{figure}
	\centering
		\includegraphics[width=0.95\linewidth]{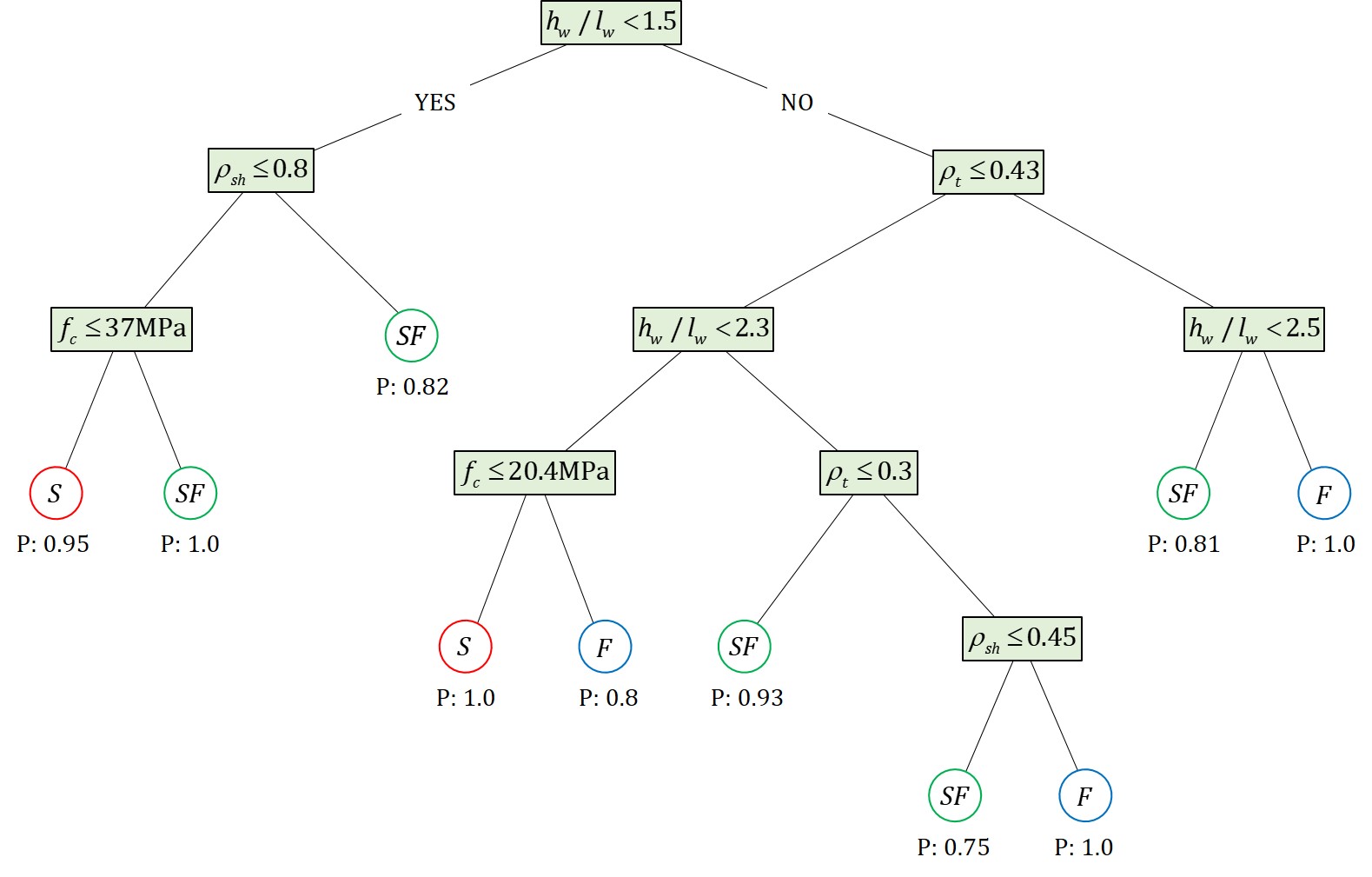}
		\caption{Proposed decision tree for seismic failure mode prediction.}
		\label{fig:proposedDT}
	\end{figure}

	Table \ref{table:examplecombos} summarizes example feature combinations along with their overall classification accuracies and Fisher Scores. Out of all possible feature combinations, the one having a high Fisher Score and high accuracy were simultaneously was selected. It is noted that Fisher Score does not necessarily guarantee high accuracy, as observed in	Table \ref{table:examplecombos}. The selected feature combination included aspect ratio ($h_w/l_w$), concrete compressive strength $(f'_c)$, transverse web reinforcement $(\rho_t)$, and boundary confinement ratio $(\rho_b)$. The DT method was then retrained using the selected four features based on 10-fold cross validation, and a proposed decision tree was finalized by evaluating 10 possible models considering their common pattern (versus picking one directly), which is presented in Fig. (\ref{fig:proposedDT}). It is noted that decision tree models were pruned to ensure physical integrity, where some decision tree models were disregarded if they are not physically correct even though they achieve higher accuracy.  The pruned decision tree is eventually proposed in Fig. (\ref{fig:proposedDT}), along with the associated probabilities (P) of making accurate predictions. For example, if a shear wall sample has an aspect ratio smaller than 1.5, includes boundary confinement ratio less than 0.8, and has concrete compressive strength smaller than 37 MPa, then the probability of this wall to be shear-controlled is 95\%. It is noted that the P value provided with the proposed decision tree being close to 1.0 implies that the predicted failure mode is most likely.

		\begin{table}
		\centering
		\caption{Example combinations of four input features.}
		\label{table:examplecombos}
		\begin{tabular}{ c c c c }
			\hline
			\textbf{Combo} & \textbf{Selected Features}               & \begin{tabular}[c]{@{}l@{}}\textbf{Fisher}\\   \textbf{Score}\end{tabular} & \begin{tabular}[c]{@{}l@{}}\textbf{DT Overall}\\   \textbf{Accuracy (\%)} \end{tabular} \\ \hline
			1     & $M/{Vl_w}$, $P/{A_gf'_c}$, $\rho_{sh}$, $\rho_{t}$ & 0.635           &77      \\ 
			2     & $M/{Vl_w}$, $P/{A_gf'_c}$, $\rho_{b}$, $\rho_{t}$  & 0.625               &79.6                                                                \\ 
			
			3     & $M/{Vl_w}$, $P/{A_gf'_c}$, $\rho_{b}$, $\rho_{sh}$      & 0.596                                                    & 77.19                                                                \\ 
			4     & $M/{Vl_w}$, $\rho_{b}$, $\rho_{sh}$,  $\rho_{t}$           & 0.554                                                    & 77.81                                                                \\ 
			5     & $M/{Vl_w}$, $P/{A_gf'_c}$,  $\rho_{t}$ , $V_n/V@M_n$  & 0.543                                                    & 69.64                                                                \\ 
			6     & $M/{Vl_w}$, $P/{A_gf'_c}$, $\rho_{sh}$, $V_n/V@M_n$ & 0.532                                                    & 73.66                                                                \\ 
			7     & $M/{Vl_w}$, $\rho_{sh}$,  $\rho_{t}$ , $V_n/V@M_n$         & 0.508                                                    & 76.50                                                                \\ 
			8     & $M/{Vl_w}$, $\rho_{b}$, $\rho_{sh}$, $V_n/V@M_n$         & 0.489                                                    & 78.92                                                                \\ 
			9     & $h_w/l_w$, $P/{A_gf'_c}$, $f'_c$, $\rho_{sh}$         & 0.437                                                    & 77.45                                                                \\ 
			10    & $h_w$/$l_w$, $P/{A_gf'_c}$, $f'_c$, $\rho_{sh}$         & 0.437                                                    & 80.26                                                                \\ 
	 	11   &    $h_w/l_w$,   $f'_c$,  $\rho_{sh}$,  $\rho_{t}$ &   0.437  & 85.29                                                                \\ 
			12    & $h_w/l_w$, $f'_c$, $\rho_{b}$, $\rho_{sh}$             & 0.437                                                    & 82.39                                                                \\ 
			13    & $h_w/l_w$, $f_yt$ / $f'_c$, $V_n/V@M_n$, X-sec & 0.185                                                    & 81.67                                                                \\ 
			14    & $h_w/l_w$, $P/{A_gf'_c}$, $\rho_{sh}$, $f_yt$ / $f'_c$    & 0.184                                                    & 82.94                                                                \\ \hline
		\end{tabular}
	\end{table}

	The proposed decision tree in Fig.\ref{table:case1_run1} was applied to the entire database  to obtain class-wise and overall accuracy. Note that the proposed decision tree is not obtained over the entire database, but is manually modified to meet the physical integrity of the problem as mentioned before.  The mean classification accuracy values are obtained as 88\%, 90\%, and 85\% for the S, the SF, and the F failure modes, respectively, with an overall accuracy of 88\%.  The results revealed that prediction accuracies were improved using four features versus using all eleven features, potentially because some features were noisy for the DT analysis. The comparisons with the ASCE 41 predictions (Table \ref{table:case1_run1}) indicate that for the S failure mode, the proposed DT model and ASCE 41 were able to make comparable predictions, whereas much higher accuracies were obtained using the proposed DT model for the remaining failure modes.

	

		\begin{table*}
			\tiny
			\centering
			\caption{Confusion matrices for the ASCE-Wall, the ASCE-Column, and the proposed DT classification model, respectively.}
					  			\label{table:confmat}

			\begin{tabular}{p{3mm}p{3mm}p{2mm}p{2mm}p{2mm}p{3mm}p{3mm} l p{3mm}p{3mm}p{2mm}p{2mm}p{2mm}p{3mm}p{3mm} l p{3mm}p{3mm}p{2mm}p{2mm}p{2mm}p{3mm}p{3mm}}
				\cline{1-7} \cline{9-15} \cline{17-23}
				\multicolumn{2}{c}{\multirow{2}{*}{\textbf{ASCE-W}}} & \multicolumn{3}{l}{\textbf{Estimated}} &                &             &           & \multicolumn{2}{c}{\multirow{2}{*}{\textbf{ASCE-C}}} & \multicolumn{3}{l}{\textbf{Estimated}} & \textbf{}      & \textbf{}   &  & \multicolumn{2}{c}{\multirow{2}{*}{\textbf{DT}}} & \multicolumn{3}{l}{\textbf{Estimated}} &                &             \\ \cline{3-7} \cline{11-15} \cline{19-23} 
				\multicolumn{2}{c}{}                                 & \textbf{S}   & \textbf{SF}    & \textbf{FC}    & \textbf{Total} & \textbf{Se} 	& \textbf{} & \multicolumn{2}{c}{}                                 & \textbf{S} & \textbf{SF} & \textbf{FC} & \textbf{Total} & \textbf{Se} &  & \multicolumn{2}{c}{}                             & \textbf{S} & \textbf{SF} & \textbf{FC} & \textbf{Total} & \textbf{Se} \\ \cline{1-7} \cline{9-15} \cline{17-23} 
				\multirow{3}{*}{\textbf{True}}    & \textbf{S}       &\bf 42  			& 14              & 2              & 58             & 72          	&           & \multirow{3}{*}{\textbf{True}}    & \textbf{S}      & \bf45           & 20            & 9            & 74             & 60            &  & \multirow{3}{*}{\textbf{True}}  & \textbf{S}     &\bf45           &1             &1             &47                &95             \\ 
				& \textbf{SF}      & 7     		& \bf44              & 50               & 101          & 43            &           &                                   & \textbf{SF}      & 4            & \bf43            & 37           & 84             & 52            &  &                                 & \textbf{SF}    &3            &\bf62            &8             &73                &85             \\
				& \textbf{FC}      & 1     		& 11               & \bf5               & 17           & 30            &           &                                   & \textbf{FC}      & 1            & 6             & \bf11           & 18             & 61            &  &                                 & \textbf{FC}    &2            &6             &\bf48            &56                &86             \\ \cline{1-7} \cline{9-15} \cline{17-23} 
				& \textbf{Total}   & 50     		& 69               & 57               & 176         &             	&           &                                   & \textbf{Total}   	  & 50           & 69            & 57           & 176            &             &  & \multicolumn{1}{l}{}            & \textbf{Total} 	  &50           &69            &57            &176               &             \\ 
				& \textbf{Pr}      & 84     		& 63               & 9               &  \textbf{OA} & \bf52            &           &                                   & \textbf{Pr}      & 90           & 62            & 19           & \bf OA             & \bf56            &  & \multicolumn{1}{l}{}            & \textbf{Pr}    &90           &90            &84            & \bf OA               &\bf 88             \\ \cline{2-7} \cline{10-15} \cline{18-23} 
				
			\end{tabular}
		\end{table*}

	To observe the performances of the proposed DT model and the  ASCE 41 criteria, the confusion matrices were provided in Table \ref{table:confmat}. The precision (Pr) was performed by dividing the total number of correctly predicted specimens by the total number of specimens that were actually associated with that failure mode, whereas the sensitivity (Se) was calculated by dividing the total number of correctly predicted specimens by the total number of specimens claimed to be associated with that failure mode. For example, based on the proposed DT model for the shear failure mode, the Pr is 45/50 = 90\%, whereas the Se is 45/47 = 95\%. 
	
	Precision and sensitivity for each failure mode were calculated based on the confusion matrix for each method. When only the precision for the shear failure mode is considered, one can say that the DT and the ASCE-C are the best classifiers. However, the sensitivity of the shear failure mode obtained using the ASCE-C criteria was 60\%, implying that even though 90\% of the shear-dominant specimens were correctly identified by ASCE-C criteria, only 60\% of them were actually true. The reason for this discrepancy is that ASCE 41 makes predictions in favor of brittle failure modes for the sake of conservatism, as discussed earlier. The ASCE 41 criteria were assessed in this paper as they are well accepted in literature; however, it should be noted that the ASCE-W and the ASCE-C need the use of only one feature and two features, respectively, whereas the proposed DT model requires four features to carry out the predictions. Therefore, from the machine learning point of view, they are not necessarily equivalent.

	\section{Summary and Conclusions}
	
	Failure modes of 176 conventional reinforced concrete shear walls tested under reverse cyclic loading were classified into three types based on experimental evidence, namely: shear (S), shear-flexure (SF), and flexural (F) failure. A physically meaningful and fully transparent machine learning-based glass-box (versus black-box) classification model was explored without compromising high accuracy. The proposed classification model ensured four important aspects, namely: high accuracy, interpretability, practicality, and physical integrity. Although, these four aspects are of similar importance in engineering, interpretability and physical integrity are commonly disregarded in ML applications in the earthquake engineering field as an increase in interpretability typically results in a decrease in model performance. This study is novel in that the trade-off between the performance and interpretability was ensured by particularly considering practicality and the physical integrity resulted in higher classification performance. 
	
	The trade-off between the interpretability and the accuracy was investigated using eight state-of-the-art machine learning methods. To provide an accurate estimate of generalization capacity of the ML methods, ten-fold cross validation was considered in all the implementations. The proposed classification model was obtained using the Decision Tree (DT) method, which was selected due to the ability of the method to provide a single, interpretable, and robust prediction model. 

As the target users are engineers, another effort was to obtain a practical and easy-to-use model that utilizes as few features (wall design parameters) as possible. Therefore, a brute-force feature selection was performed by only considering the DT, based on Fisher score, SVM-RFE, as well as engineering judgment, which reduced the original features to 4 features. The compressive strength of concrete ($f'_c$), wall aspect ratio ($h_w/l_w$), transverse boundary ($\rho_{b}$),  and web reinforcement ($\rho_{t}$) ratio were found to be the most significant features in failure mode classification. The ability of the proposed DT model to predict the failure modes was validated with actual failure modes. Although the DT achieved high prediction accuracy (above 85\%) for all three failure modes, even higher accuracies were obtained using other methods which provide black-box type classification algorithms (tools). The reason for specifically selecting the DT is that it provides a more interpretable classification model than the others (e.g. the SVM with RBF kernel, the GPC, and the RF), each of which produces impenetrable black-box type classification models.
	
	Accurate predictions of wall failure modes will enable reliable and efficient seismic risk assessment in reinforced concrete shear wall buildings. The current finite element modeling approaches (e.g. Modified Compression Theory, Fixed-Strut- Angle Finite Element model) are cumbersome and require certain modeling assumptions, yet are commonly used to predict response and failure of reinforced concrete structural elements; however, the proposed classification model will be advantageous as it by-passes modeling uncertainties and computational costs of the analytical FEM models. The outcomes of the proposed classification model may also be used as preliminary predictions to achieve more accurate modeling in a detailed risk/performance assessment. For example, modeling of shear behavior can be elaborated in the FEM model of a shear wall if the proposed classifier revealed that the wall is shear-controlled. It is also worth noting that the proposed classification model is data-driven, meaning that the outcomes depend on the dataset used in this study although the physical behavior of the shear walls is strongly considered.

	In the context of the type of  ML methods, the linear methods have always been very popular than the nonlinear methods  because of their high interpretability. However, it should be noted that the ability of the linear methods is limited when the dataset follow a nonlinear distribution, hence, yielding a  poor classification performance. Therefore, there is a trade-off between the classification performance and the interpretability of the decision functions. The aim of this paper is to explore the feasibility of the machine learning algorithms in wall failure mode classification in terms of both the performance and the practicality. 
	
%
%
\clearpage
\bibliographystyle{spphys}
\bibliography{mainZD}
%
%

\end{document}